\documentclass[twocolumn,english,aps,prb,showpacs,showkeys]{revtex4}
\usepackage[T1]{fontenc}
\usepackage[latin1]{inputenc}
\usepackage{amsmath}
\usepackage{color}
\usepackage{graphicx}
\usepackage{amssymb}

\makeatletter

\usepackage{epsfig}

\usepackage{babel}
\makeatother
\begin{document}

\title{Spin- and Charge Excitations of the Triangular Hubbard-Model: a FLEX-Study}

\author{Marcus Renner}

\email{m.renner@tu-braunschweig.de}

\author{Wolfram Brenig}

\affiliation{Technische Universit\"{a}t Braunschweig, Institut f\"{u}r Theoretische
Physik, Mendelssohnstr. 3, 38106 Braunschweig}

\date{\today{}}

\begin{abstract}
A study of the quasi-particle excitations and spin fluctuations in the
one-band Hubbard-model on the triangular lattice with
nearest- and next-nearest-neighbor hopping is presented. 
Using the fluctuation-exchange-approximation (FLEX) results for the
quasi-particle dispersion and life-time, the Fermi surface, and the
static spin structure factor will be discussed for a wide range of
dopings and as a function of the Coulomb correlation strength $U$.
It is shown that the renormalization of the spin- and charge-dynamics
is sensitive to the interplay between van Hove singularity-effects
and the nesting, which is influenced by the next-nearest-neighbor hopping. For
all dopings investigated, the energy-dependence of the quasi-particle
life time is found to be of conventional Fermi-liquid nature. At intermediate
correlation strength the static structure factor is strongly doping
dependent, with a large commensurate peak at the $K$-point for 1.35
electrons per site and weak, incommensurate intensities occuring at
lower electron densities. 
The relevance of this model to the
recently discovered cobaltates Na$_{x}$CoO$_{2}\cdot y$H$_{2}$O
will be discussed.
\end{abstract}



\maketitle

\section{Introduction}

Recent discovery of superconductivity in the cobaltate Na$_{x}$CoO$_{2}\cdot y$H$_{2}$O,
$x\approx 0.35$ with a critical temperature of $T_{c}\sim 5K$ \cite{Takada2003}
has attracted enormous attention to this material, stimulating a large
number of experimental 
\cite{Lorenz2003,Schaak2003,Sakurai2003,Hasan2003}
and theoretical investigations 
\cite{Baskaran2003,Kumar2003,Wang2003,Honerkamp2003,Singh2003,Tanaka2003,Ogata2003,Wang2003b,
Li2003,Zou2003,Kunes2003}.
Na$_{x}$CoO$_{2}\cdot y$H$_{2}$O is of space group P6$_{3}$/mmc,
representing a hexagonal layered structure which consists of two-dimensional
(2D) CoO$_{2}$ conduction planes, separated by H$_{2}$O and Na$^{+}$
ions. The material is derived from its parent compound NaCo$_{2}$O$_{4}$,
equivalent to $x$=0.5, by simultaneous intercalation of H$_{2}$O
and deintercalation of Na$^{+}$ ions. This process nearly doubles
the c-axis lattice-constant, strongly enhancing the 2D character
of the compound. The Co ions form a \emph{triangular} lattice and
are coordinated by edge-sharing oxygen-octahedra which display a shift
of the oxygen atoms along the $\left(1,1,1\right)$-direction. In
this environment the low-lying $t_{2g}$-manifold splits into
a non-degenerate $A_{1g}$ level and an $E_{g}$-doublet, leaving
Co$^{4+}$ in a low-spin, $S$=1/2 state, with a single electron in
the $A_{1g}$ level while Co$^{3+}$ has a filled $A_{1g}$ orbital
\cite{Singh2000}. In turn, $x$ can be viewed as \emph{electron
doping} away from half-filling of a non-degenerate one-band model
on a triangular lattice, with Na$_{x}$CoO$_{2}\cdot y$H$_{2}$O,
$x\approx 0.35$ refering to an electron density of $n\approx 1.35$
per site. 

At present, the electronic structure of the cobaltates is an open
issue. Yet, there is a tempting similarity to the high-temperature
superconducting cuprates, where however $d$-holes are arranged
on a 2D \emph{square}-lattice. Early on, the role of electron correlations
in the cobaltates has been of interest. LDA results in an $A_{1g}$
bandwidth of $W\approx $1-1.4eV, while the $d$-shell Coulomb-correlation
strength $U$ is of order 5-8eV typically \cite{Singh2000}. Motivated
by this, resonating-valence-bond (RVB)\cite{Anderson73,Fazekas74} 
scenarios have been pursued by several researchers 
\cite{Baskaran2003,Wang2003,Kumar2003,Ogata2003, Li2003}.
In addition, weak-coupling
calculations, using renormalization-group (RG) methods have also been
performed \cite{Honerkamp2003}. In this paper we apply an alternative
method, suitable for intermediate correlation strength, namely the
fluctuation-exchange-approximation (FLEX) \cite{Bickers89,Altmann2000}
and investigate the spin- and charge-dynamics in the triangular Hubbard-model.
In passing, we note that such analysis may also shed light onto the
organic superconductors $\kappa $-(ET)$_{2}$X \cite{etrev}. These
have been suggested to realize the Hubbard-model on an \emph{anisotropic}
triangular lattice \cite{etrev}.

The paper is organized as follows: in section \ref{method} we discuss
the model and briefly review the FLEX method. In section \ref{results}
results for the renormalized dispersions, Fermi surfaces, 
the static spin structure factor, 
and the frequency dependence of the selfenergy are
discussed for several dopings.

\section{Model and Method\label{method}}

Our subsequent analysis is focussed on the normal state properties of
the one-band Hubbard-model on the triangular lattice

\begin{equation}
	H = - \sum _{ij}t_{ij}(c_{i}^{\dagger }c_{j}^{\phantom \dagger }+h.c.)
	    +U\sum _{i}n_{i\uparrow }n_{i\downarrow }\, .
\label{HubMod}
\end{equation}

\begin{figure}[t]
\begin{center}\includegraphics[  width=1.0\columnwidth,
  keepaspectratio]{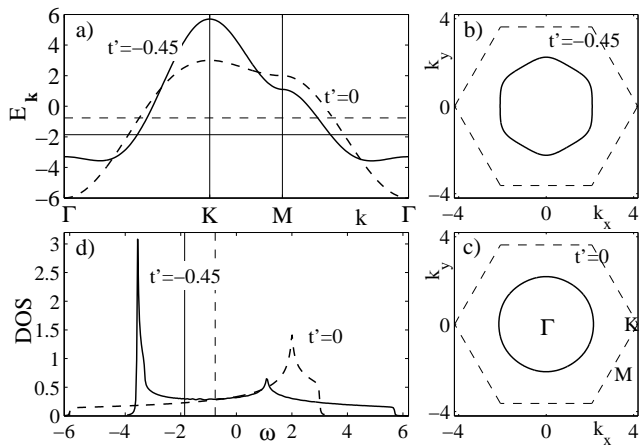}\end{center}

\caption{\label{gridfs} a) dispersion, b), c) Fermi surface, and d) density
of states for the thight-binding energy of eqn. (\ref{eq:w1}) for
$t'=0$ versus $t'\neq 0$. In a) {[}d){]} horizontal{[}vertical{]}
solid(dashed) lines refer to chemical potentials $\mu $ at $n=0.65$.
In b) and c) dahed hexagons label Brillouin zone of triangular lattice.}
\end{figure}

Due to the lack of electron-hole symmetry, 
a choice has to be made
to fix the sign of the the hopping matrix elements $t_{ij}$ and the Coulomb
repulsion $U$. 
Here we fix these parameters to model the electronic structure
of Na$_{x}$CoO$_{2}\cdot y$H$_{2}$O which we assume to be similar to the one of
the parent compound. LDA calculations on NaCo$_{2}$O$_{4}$ \cite{Singh2000}
place the Fermi energy in a CEF-split $t_{2g}$ manifold of Co $d$-bands
with the states near the Fermi energy $E_{F}$ to be of $a_{1g}$
symmetry. Due to a weak unit-cell doubling a 'bilayer'-type splitting
exists which we will neglected henceforth. LDA finds the maximum of
the $a_{1g}$-$t_{2g}$ bands at the $\Gamma $-point. Focussing at
first on the nearest-neighbor(n.n.)-hopping $t=t_{\langle ij\rangle }$,
and using an electron-language this implies that $t>0$ in eqn. (\ref{HubMod}).
Coulomb-correlations however, suppress double occupancy of the $a_{1g}$-$t_{2g}$
level by holes. Therefore, for the remainder of this paper we switch
to the hole-language, implying $t>0$ and $U>0$. The corresponding
tight-binding band and bare DOS, as well as the Fermi surface (FS)
and chemical potential for $n\approx 0.65$ are shown in Fig.~\ref{gridfs}.
In the remainder of this paper all energies will be expressed in units of $t$.

Next, it is important to realize that a dispersion with n.n.-hopping
only is \emph{qualitatively} different from the LDA. First, the latter
exhibits a pronounced dip at the $\Gamma $-point. Second, the LDA
FS for the parent compound NaCo$_{2}$O$_{4}$, i.e. at $n$=0.5,
is notably hexagonal, while the tight-binding FS using n.n.-hopping
only is practically circular at this doping. These facts signal
a sizeable next-nearest-neighbor(n.n.n.) hopping 
$t'=t_{\langle \langle ij\rangle \rangle }$
in the cobaltates. Fitting the corresponding dispersion

\begin{alignat}{1}
\epsilon _{\mathbf{k}} & =-2t(\cos \left(k_{x}\right)+2\cos (k_{x}/2)\cos (\sqrt{3}k_{y}/2))\nonumber \\
 & -2t'(2\cos (3k_{x}/2)\cos (\sqrt{3}k_{y}/2)+\cos (\sqrt{3}k_{y}))\label{eq:w1}
\end{alignat}
to the LDA along the $\Gamma $-$M$ direction we find $t'\approx -0.45$.
As can be read of from Fig.~\ref{gridfs} this does not only lead
to the required dip in the dispersion and the hexagonal FS shape,
but also to a signature of van Hove singularities very different from
the case $t'=0$. For $t'=0$ the bandwidth is $W=9$, while for
$t'=-0.45$ it is slightly increased to $9.26$.
The role of n.n.n. hopping has also been emphasized in a recent
search for possible $f$-wave SC \cite{Ikeda2003} using third order pertubation
in $U/t$ for the gap equation, which is rather different from the
approach presented here.

To study the model (\ref{HubMod},\ref{eq:w1}) we employ the diagrammatic
FLEX approximation\cite{Bickers89}. The FLEX is conserving in the
sense of Kadanoff and Baym \cite{Baym62}. It sums all particle-hole(particle)
ladder-graphs for the generating functional selfconsistently and it
is believed to apply up to intermediate correlation strenght $U/t\sim 1$.
The FLEX equations for the bare(renormalized spin- and charge-) susceptibilities
$\chi ^{0}$($\chi ^{s,c}$), the one-particle Greens function(selfenergy)
$G$($\Sigma $), and the effective interaction $V$ read

\begin{eqnarray}
\chi _{\mathbf{q}}^{0}\left(\nu _{m}\right) & = & -\frac{T}{N}\, \sum _{\mathbf{k},n}G_{\mathbf{k}+\mathbf{q}}\left(\omega _{n}+\nu _{m}\right)G_{\mathbf{k}}\left(\omega _{n}\right)\label{ph-bubble}\\
\chi _{\mathbf{q}}^{s,c}\left(\nu _{m}\right) & = & \chi _{\mathbf{q}}^{0}\left(\nu _{m}\right)\left[1\mp U\chi _{\mathbf{q}}^{0}\left(\nu _{m}\right)\right]^{-1}\label{chi_sc}\\
V_{\mathbf{q}}\left(\nu _{m}\right) & = & U^{2}(\frac{3}{2}\chi _{\mathbf{q}}^{s}\left(\nu _{m}\right)+\frac{1}{2}\chi _{\mathbf{q}}^{c}\left(\nu _{m}\right)-\chi _{\mathbf{q}}^{0}\left(\nu _{m}\right))\label{pot}\\
\Sigma _{\mathbf{k}}\left(\omega _{n}\right) & = & \frac{T}{N}\, \sum _{\mathbf{k}',n'}V_{\mathbf{k}-\mathbf{k}'}\left(\omega _{n}-\omega _{n'}\right)G_{\mathbf{k}'}\left(\omega _{n'}\right)\label{selfen}
\end{eqnarray}
where $\omega _{n}=i\pi T(2n+1)$ and $\nu _{m}=i\pi T2m$. Eqns.
(\ref{ph-bubble}-\ref{selfen}) have to be solved selfconsistently
taking into account that 
$G_{\mathbf{k}}^{-1}\left(\omega _{n}\right)=\left[\omega _{n}-\epsilon _{\mathbf{k}}+\mu -\Sigma _{\mathbf{k}}\left(\omega _{n}\right)\right]^{-1}$.
We compute the Matsubara summations using the 'almost real contour'
technique of ref. \cite{Schmalian96}. I.e., contour integrals are
performed with a \emph{finite} shift $i\gamma $ ($0<\gamma <iT\pi /2$)
into the upper half plane. All final results are analytical continued
from $\omega +i\gamma $ onto the real axis $\omega +i0^{+}$ by Pad\'{e}
approximation. The following results are based on FLEX solutions using
a lattice of 64x64 sites with 4096 equidistant $\omega $-points in
the energy range of $[-30,30]$. The temperature has been kept at
$T=0.05$ .

\section{Results and Discussion\label{results}}

\begin{figure}[t]
\begin{center}\includegraphics[  width=0.90\columnwidth,
  keepaspectratio]{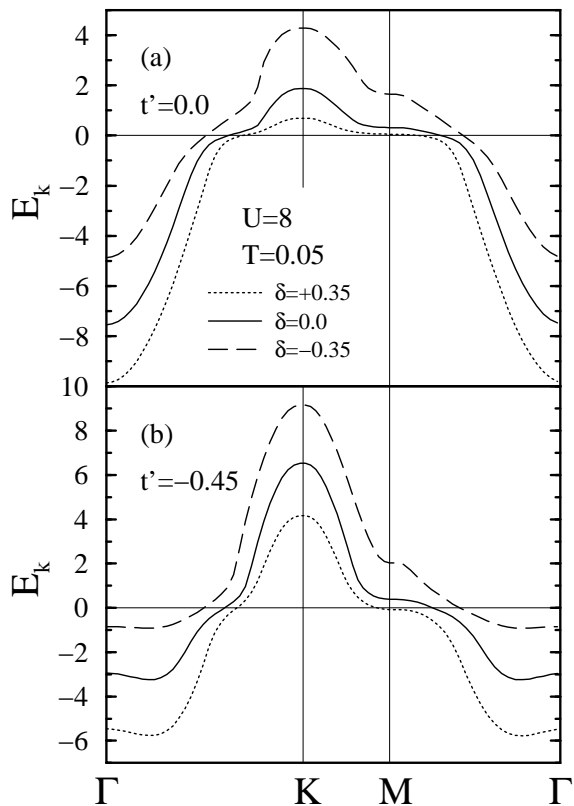}\end{center}
  
\caption{\label{disp}
Quasi-particle
dispersion $E_{\mathbf{k}}$ along $\Gamma $-$K$-$M$-$\Gamma $ for (a) $t'=$0
and (b) $t'=-0.45$ and for various doping levels at $U=8$ and $T=0.05$.}
\end{figure}

We start the discussion of our results with the quasi-particle dispersion
$E_{\mathbf{k}}$ which is determined by 
$E_{\mathbf{k}}-\epsilon _{\mathbf{k}}+\mu -\Sigma _{\mathbf{k}}(E_{\mathbf{k}})=0$.
This dispersion is shown in Fig.~\ref{disp}(a) and (b) at $U/W\sim 1$
for $t'=0$ and $-0.45$ for various doping levels $\delta $,
where $\delta =n-1$ within the hole-picture,
i.e. Na$_{x}$CoO$_{2}\cdot y$H$_{2}$O with $x\approx 0.35$ corresponds
to $\delta \approx -0.35$. As compared to Fig.~\ref{gridfs} (a)
a pronounced mass-enhancement of order unity at the FS crossings is
clearly visible in Fig.~\ref{disp} (a) and (b). This enhancement
is due to low-energy spin fluctuations which are present in 
$\chi _{\mathbf{q}}^{s}\left(\omega \right)$.
In addition the overall shape of the dispersion is strongly renormalized.
At large distances from the FS however, and due to strong quasi-particle
scattering, the notion of the quasi-particle pole $E_{\mathbf{k}}$
may be of less relevance. 

At $U=0$, $t'=0$, and
for $\delta = +0.5$ the Fermi energy coincides with the van
Hove singularity related to a flattening of the bare dispersion observable
in Fig.~\ref{gridfs} (a) along $\Gamma $-$M$-$K$. Fig.~\ref{disp}
(a) shows a shift of this flattening at finite $U$. Eg. for
$U=8$ we find the Fermi energy to coincide with the remnants of the
van Hove singularity only at $\delta \approx +0.35$. This is rather
remote from the SC's doping level. 
In ref. \cite{Honerkamp2003} Fermi surface pinning of the van Hove 
singularity has also been investigated, however at fixed chemical potential.
For $t'\neq 0$ the situation
is very different since \emph{two} van Hove singularities exist in
the bare DOS. Remnants of these remain present also at finite $U$.
At $t'=-0.45$ and $U=8$ Fig.~\ref{disp} (b) shows that for the SC's
doping level of $\delta \approx -0.35$ the FS is rather close to
the lower one of the van Hove singularities - which is absent at $t'=0$.
Finally we note from Fig.~\ref{disp} (a), that for $t'=0$ the relative
mass-enhancement at the FS along directions unrelated to the van Hove
singularity, e.g. $\Gamma $-$K$ is largest if the van Hove singularity
is closest to the FS. For $t'=-0.45$ Fig.~\ref{disp} (b) does not
show this effect.

\begin{figure}
\begin{center}\includegraphics[  width=0.90\columnwidth,
  keepaspectratio]{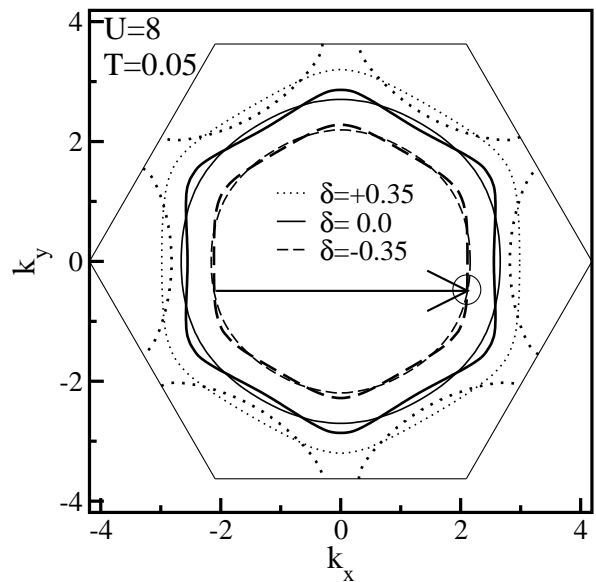}\end{center}

\caption{\textbf{\label{fs}} 
Fermi surfaces for $t'=0$ (thin lines) versus $t'=-0.45$ (thick lines)
at $U=8$ and $T=0.05$ for various $\delta $. Outer 
hexagon refers to BZ. Arrow denotes a commensurate wave vector 
${\bf K} =(\frac{4\pi }{3},0)$
with a circle labeling the half width of $\chi _{\mathbf{q}}^{s}$
at $\mathbf{q}={\bf K}$ at $T=0.05$, $U=8$, and $t'=0$.}
\end{figure}

In Fig.~\ref{fs} the FS of the $t-$ and $t-t'-U$ models are compared
for $\delta =0, \pm 0.35$. As compared to the non-interacting
case at $U=0$ there are no indications of sizeable, correlation-induced
shape-changes of the FS. 
The FS at finite and negative $t'=-0.45$
is hexagonally shaped 
for $\delta \le 0$ down to the SC's doping level.
This is consistent with the
LDA results of ref. \cite{Singh2003}. In contrast, the FS at $t'=0$
evolves towards a more circular shape as $\delta =-0.35$ is reached.
Nesting of the flat regions of the FS with
the wave-vector $\mathbf{Q}\left(\delta \right)$ indicated in 
Fig.~\ref{fs} can enhance the static spin structure factor at 
$\mathbf{Q}$$\left(\delta \right)$.
Due to the shape of the FS this enhancement is more pronounced for
$t'=-0.45$ at $\delta\le 0$. Interestingly, at the SC's doping level the nesting-vector
$\mathbf{Q}$ is \emph{commensurate}, i.e. 
$\mathbf{Q}\left(\delta \approx -0.35\right) = {\bf K} 
= \left(\frac{4\pi }{3},0\right).$

\begin{figure}
\begin{center}\includegraphics[  width=0.90\columnwidth,keepaspectratio]{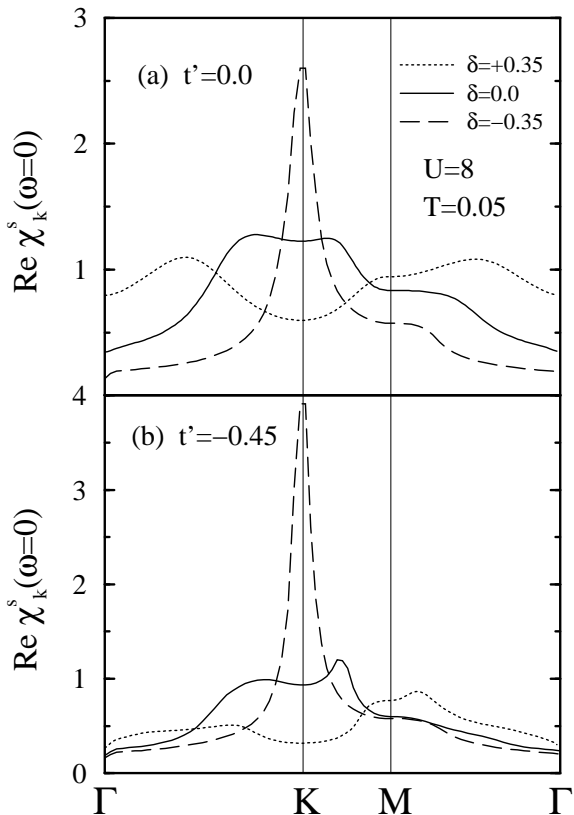}\end{center}

\caption{\label{tsf1}
Doping dependence of the static spin structure factor
$\mathrm{Re}\, \chi _{\mathbf{k}}^{s}(\omega =0)$
along the path $\Gamma $-$K$-$M$-$\Gamma $ in the BZ
for (a) $t'=0$ and (b) $t'=-0.45$.
Notice that only at $\delta =-0.35$ the commensurate peak at 
${\bf K}$ is observed for large $U$.}
\end{figure}

\begin{figure}[t]
\begin{center}\includegraphics[  width=0.90\columnwidth]{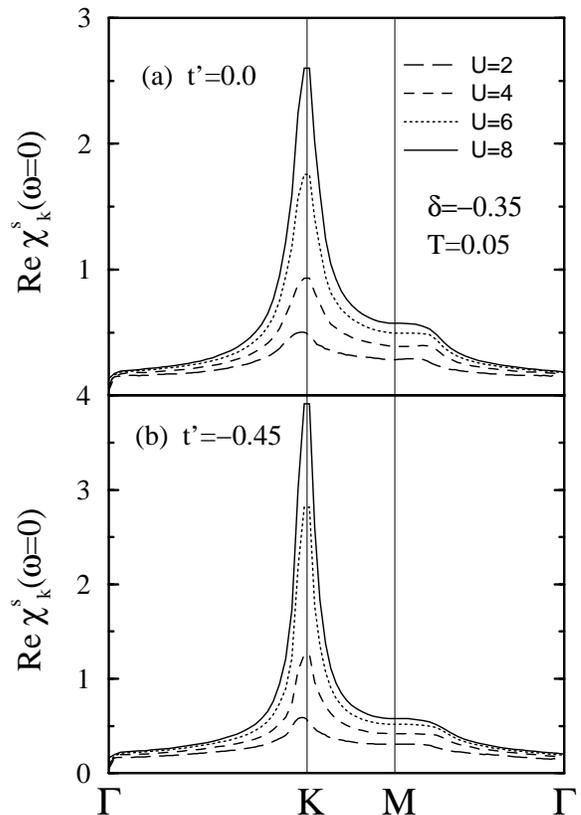}\end{center}

\caption{\label{tsf2}
Dependence
of the static spin structure factor 
$\mathrm{Re}\, \chi _{\mathbf{k}}^{s}(\omega =0)$
on the correlation strength $U$
along the path $\Gamma $-$K$-$M$-$\Gamma $ in the BZ
for (a) $t'=0$ and (b) $t'=-0.45$.}
\end{figure}

Now, we discuss the evolution of the static spin structure factor
$\textrm{Re}\left[\chi _{\mathbf{k}}^{s}(\omega =0)\right]$ with
dopings $\delta $ and correlation strength $U$ along the path $\Gamma $-$K$-$M$-$\Gamma $
in momentum space, contrasting the case of $t'=0$ versus $t'=-0.45$.
Fig.~\ref{tsf1} displays the doping dependence at $T=0.05$ and $U=8$.
For $\delta =+0.35$ this figure shows the static structure factor
to be only moderately enhanced whithin a rather broad range of incommensurate
momenta. 
For $t'=0$ these momenta are located mainly along $\Gamma-M$ and partially along $\Gamma-K$.
The zone-folded approximate nesting vector is located slightly off the $M$-Point.
For $t'=-0.45$ and $\delta =+0.35$ enhanced spin fluctuations are
found primarily along the direction $\Gamma $-$M$ with 
$\textrm{Re}\left[\chi _{\mathbf{k}}^{s}(\omega =0)\right]$ rather small within
the remaining $\mathbf{k}$-space. As the doping is reduced the maxima
in $\textrm{Re}\left[\chi _{\mathbf{k}}^{s}(\omega =0)\right]$ move
towards the $K$-point and and develop into a sharp and \emph{commensurate}
peak at $\mathbf{Q}(\delta) = {\bf K}$
as the SC's doping level of $\delta =-0.35$
is reached. 
The shoulder at $M$ in Fig. \ref{tsf2} is due to the overlap 
with the remaining peaks at symmetry equivalent positions. 
Nearly all incommensurate spin fluctuations are suppressed.
The half width of the commensurate spin fluctuation peak for 
$\delta =-0.35$ and $t'=0$ is scetched by the circle on the right hand side 
of the inner FS in Fig.~\ref{fs}. 
This circle is also a measure for the degree of nesting of the quasi-parallel
regions of the Fermi surface.
At $\delta =-0.35$, close to the SC's doping level, the commensurate peak is
$\sim $60\% larger for $t'=-0.45$ than for $t'=0$. 
At $\delta =+0.35$ however, where the nesting properties of the
$t-$ and $t-t'-U$ models are rather reversed, the spin response is less supressed 
for $t'=0$.
 
In Fig.~\ref{tsf2} (a) and (b) the dependence of the static spin
structure factor on the correlation strength $U$ is displayed at
the SC's approximate doping concentration for $t'=0$ and $t'=-0.45$
along a path in the BZ identical to that of Fig.~\ref{tsf1}. Evidently,
increasing $U$ at this doping level, predominantly increases 
$\textrm{Re}\left[\chi _{\mathbf{k}}^{s}(\omega =0)\right]$
at the commensurate momentum ${\bf K}$. Additionally, one realizes that this
increase is non-linear in $U$, with a rather sizeable growth setting
in for $U\gtrsim 4$. Finally, and similar to the doping dependence,
the overall size and sensitivity of the spin fluctuations to the correlation
strength is larger at $t'=-0.45$ than at $t'=0$. In contrast to
the situation at $\delta =-0.35$ we have found the incommensurate 
spin fluctuation peaks at $\delta =+0.35$ to be enhanced only weakly 
with $U$ increasing up to $U=8$.

\begin{figure}[tb]
\begin{center}\includegraphics[  width=0.90\columnwidth,
  keepaspectratio]{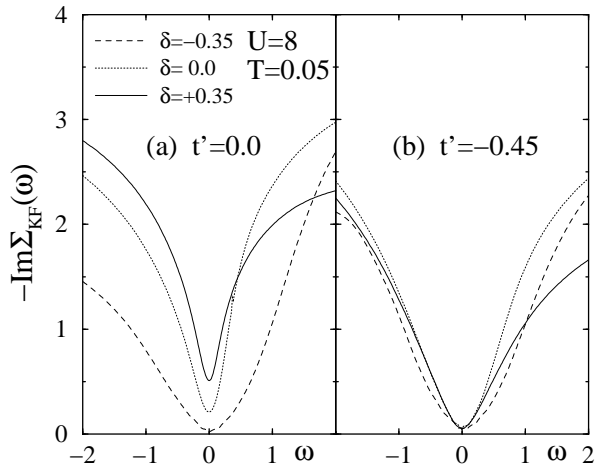}\end{center}

\caption{\textbf{\label{ims_w}}
Frequency dependence of $-\mathrm{Im}\Sigma _{\mathbf{k}}\left(\omega \right)$
near the FS in direction $\Gamma \rightarrow K$ 
for (a) $t'=0$ and (b) $t'=-0.45$
and $\delta=0, \pm 0.35$
(location of momenta ${\bf k}$ in units of ${\bf K}$ for
(a): $\delta =-0.35$: 33/64; $\delta =0.0$: 42/64; $\delta =+0.35$: 45/64
and
(b): $\delta =-0.35$: 33/64; $\delta =0.0$: 39/64; $\delta =+0.35$: 42/64
).}
\end{figure}

Fig.~\ref{ims_w} shows the low-temperature frequency dependence
of the imaginary part of the quasi-particle selfenergy 
Im$\Sigma _{\mathbf{k}\in FS}(\omega )$
for $\delta =0$, and $\pm 0.35$. The lattice momenta are chosen
as close as possible to the FS. We find the selfenergy to be nearly
isotropic along the FS with only a rather weak maximum occuring in
the direction of the commensurate spin fluctuations.
Near the Fermi energy and for \emph{all} dopings
shown, the selfenergy is clearly proportional to $\omega ^{2}$ at
low-energies which is indicative of normal Fermi-liquid behavior.
This is in sharp contrast to FLEX analysis of the Hubbard-model on
the \emph{square lattice}, close to half filling. There one typically
finds 'marginal' Fermi-liquid behavior with 
Im$\Sigma _{\mathbf{k}\in FS}(\omega )\propto \omega $
over a wide range of frequencies \cite{Altmann2000, Wermbter97}. Therefore, along
this line one is tempted to conclude, that the normal state of the
cobaltates is more of conventional metallic nature than in the cuprates.
This is even more so, if one realizes from Fig.~\ref{ims_w} that
the quasi-particle scattering rate displays its smallest curvature
at the SC's doping level, which implies the quasi-particles to be
rather well defined there. At larger values of $\delta $, proximity
of the FS to the van Hove singularities enhance both, the absolute
value of Im$\Sigma _{\mathbf{k}\in FS}(\omega )$ as well as the curvature.
This effect is more pronounced for $t'=0$.

In summary we have investigated the one-band Hubbard-model on a triangular
lattice with nearest- and next-nearest-neighbor hopping using the
normal state version of the FLEX approximation. Adapting this model
to the cobaltates we have shown that a sizeable next-nearest-neighbor
hopping is necessary to describe their electronic structure. Over
a substantial range of dopings which we have investigated we find
a significant Fermi surface mass-enhancement of order unity. This
is due to spin fluctuations and the additional proximity of the Fermi surface
to van Hove singularities at particular doping levels. In contrast
to the Hubbard-model on the square-lattice we have found the quasi-particle
scattering rate to display a conventional Fermi-liquid type of energy-dependence.
Finally we have shown that the static spin structure factor exhibts
a large commensurate peak at those doping levels attributed to the
superconducting cobaltate systems. This response was found to be significantly
enhanced by next-nearest-neighbor hopping.

\end{document}